\documentstyle[aps,prl,twocolumn,epsf,rotate,graphics]{revtex}

\begin{document}
\twocolumn[\hsize\textwidth\columnwidth\hsize\csname
@twocolumnfalse\endcsname

\title{Theoretical models of ferromagnetic III-V semiconductors}

\draft

\author{T. Jungwirth$^{1,2}$, Jairo Sinova$^{2}$, J. Ku\v{c}era$^{1}$ and A. H. MacDonald$^{2}$ 
}
\address{
$^{1}$Institute of Physics ASCR, Cukrovarnick\'a 10,  
162 53 Praha 6, Czech Republic}
\address{
$^{2}$Department of Physics, The University of Texas, Austin, TX 78712}
\date{\today}
\maketitle

\begin{abstract}

Recent materials research has advanced the maximum 
ferromagnetic transition temperature in semiconductors containing 
magnetic elements toward room temperature.  Reaching this goal would make
information technology applications of these materials likely.
In this article we briefly review the status of work over the past five years
which has attempted to achieve a theoretical understanding of these complex magnetic systems. 
The basic microscopic origins of ferromagnetism in the (III,Mn)V compounds
that have the highest transition temperatures appear to be well understood,
and efficient computation methods have been developed which are able to 
model their magnetic, transport, and optical properties.  However many
questions remain.

\end{abstract}


\vskip2pc]

It is hoped that the 
emerging field of semiconductor spintronics, which studies 
the controlled flow of charge {\em and spin} in a semiconductor, 
will lead to the development of new non-volatile, high-density, and high-speed 
information technologies.  An important milestone in this field was 
the discovery five years ago of ferromagnetism in  Mn-doped,
p-type  GaAs  observed  at temperatures  in  excess  of  100  Kelvin
\cite{ohnoscience98}.  Ferromagnetism at room temperature 
with full participation of itinerant carriers 
would be a major breakthrough in semiconductor spintronics.
With this aim, intensive material research is currently in progress on
transition metal doped III-V semiconductors.
In this brief review we present a snap shot of 
the theoretical part of this endeavor,
which has progressed hand in hand with experimental developments.

\begin{figure}[h]
\epsfxsize=3.15in
\epsffile{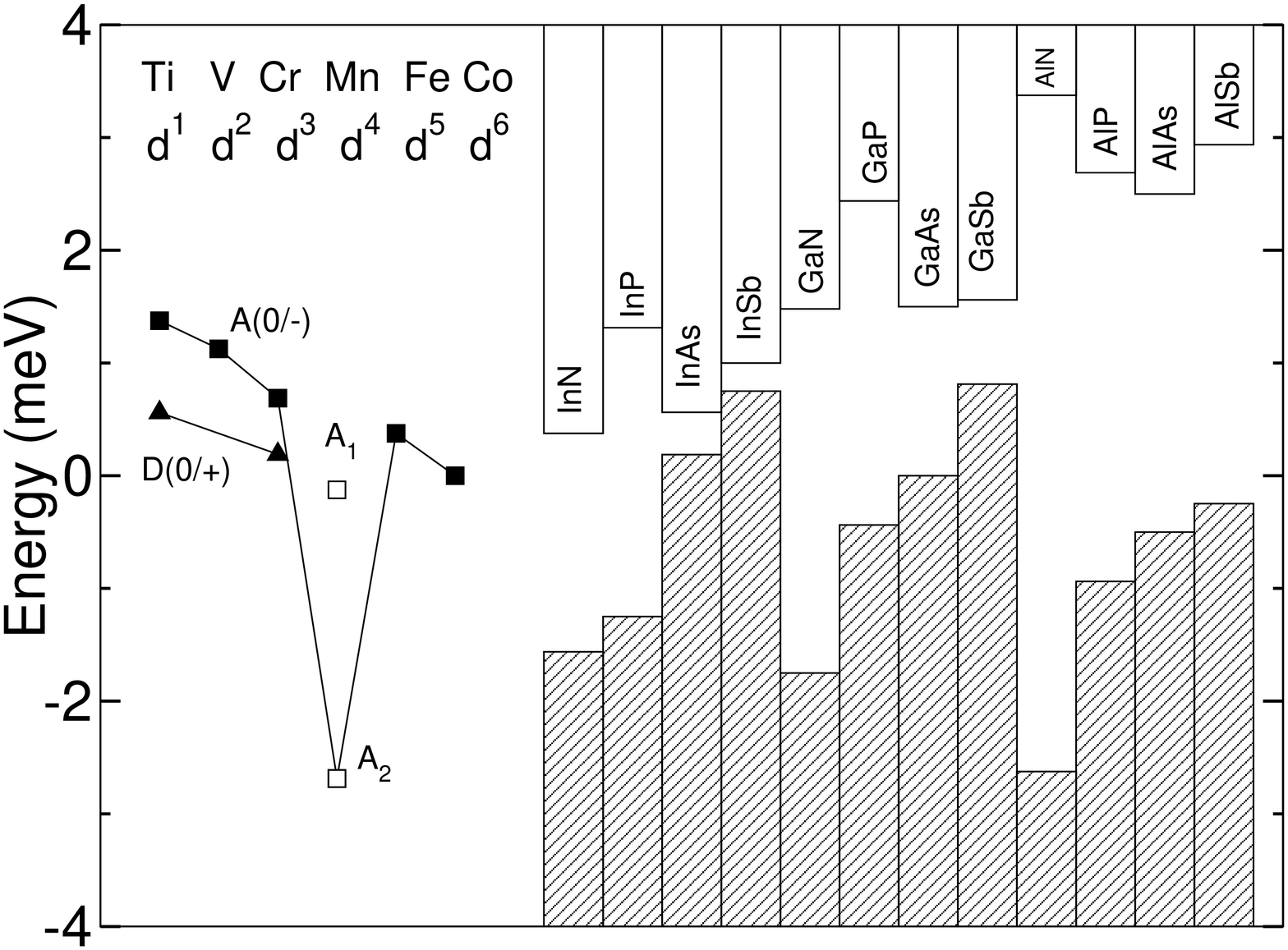}
\caption{Approximate positions of transition-metal donor (D(0/+)) and
acceptor (A(0/-)) levels in III-V compounds \protect\cite{blinowski0201012}.
Acceptor state labeled by A$_1$ and the deep A$_2$ level were obtained from 
spin-resonance experiments in GaP:Mn \protect\cite{kreisslprb96}
and from photoemission in (Ga,Mn)As \protect\cite{okabayashiprb98}, 
respectively.}
\label{dietl}
\end{figure}

A qualitative picture of the electronic structure of III-V
diluted magnetic semiconductors (DMSs) was proposed by Dietl and coworkers
\cite{dietlsst02,blinowski0201012}.  Their model is based on the internal
reference rule \cite{langerprb88} which  states  that energy  levels
derived from  the d-shell of the magnetic ion are constant across 
semiconductor compound families with properly aligned bands.
The application of this rule to III-V  materials is illustrated
in Fig~\ref{dietl} for ionized magnetic impurities substituted on cation
sites \cite{dietlsst02,blinowski0201012,kreisslprb96,okabayashiprb98}. 
The position of the A$_2$ level
for Mn in GaAs, deep in the valence band,
 suggests that Mn in GaAs is 2+, that its d-shell is occupied
by five electrons, and that incorporation of Mn in this and many other 
(III,Mn)V compounds will result a large density of valence band holes that  
can mediate ferromagnetic coupling between the  $S=5/2$ Mn local  moments.
For low carrier densities, these valence band holes will be bound 
to the Mn ion, leading
to shallow acceptor levels.  This  model of carrier-induced
ferromagnetism is now fairly well established for (Ga,Mn)As as a result of
electron paramagnetic resonance experiments \cite{szczytkoprb01}, x-ray
magnetic circular dichroism  measurements 
\cite{beschotenprl99,ohldagapl00}, and 
magneto-transport data \cite{ohnojmmm99,baxterprb02,edmonds0205517,gallagher}.

The acceptor impurity levels of Fe and Co
\cite{dietlsst02,blinowski0201012,langerprb88,kreisslprb96}
 are unlikely to lead to high valence band hole concentrations
in arsenides or antimonides,  as shown in
Fig.~\ref{dietl}.   The possible  coexistence of  acceptor  and neutral
magnetic  impurities  suggests  that a  ferromagnetic  double-exchange
mechanism can  dominate in  this case \cite{akaiprl98}.
The scenario, however, has not yet  been confirmed experimentally.
For nitrides and phosphides the ionization energy of the acceptor levels is
even higher and the A$_2$ Mn impurity level is not
deep in the valence band, making it likely that d-charge fluctuations 
cannot be neglected.  
Experimentally, ferromagnetism with critical temperatures 
exceeding room-temperature
has been claimed in (Ga,Mn)N \cite{reedapl01,sasakiapl02} on the basis of 
hysteresis in magnetization measurements, however, the electronic configuration
of Mn impurity in this compound has not yet been conclusively established,
and these claims have not yet been widely confirmed. 
The observation of the
$p$-type conduction achieved in some samples  \cite{sonoda0205560}
seems to suggest
that the simple picture of A$_1$ Mn acceptor
level according to Fig.~\ref{dietl} 
may not apply in this case.  

Guided by the complex phenomenology very 
briefly discussed above and by available
experimental data on  III-V DMSs,  several theoretical groups  have performed
detailed   studies  of   these  new   ferromagnetic   materials  using
both microscopic and effective Hamiltonian  approaches.  The results
of some of this theoretical research are
reviewed in the following paragraphs.

Ab-initio  theories are  an invaluable tool for studying the microscopic
origins  of ferromagnetism  and for predicting  electronic,  magnetic, and
structural  ground-state properties of  new compounds.   The local density
approximation  (LDA)  combined   with  disorder-averaging  coherent-potential
approximation (CPA)  was used \cite{satosst02}  to study GaAs  and GaN
DMSs  doped  with  V, Cr,  Mn,  Fe,  Co,  and Ni.   A ferromagnetic
ground-state was predicted for V, Cr,  and Mn doping while Fe, Co, and
Ni  impurities were predicted to lead to  a spin-glass  state.
The structural  properties of
(Ga,Mn)As  have  been calculated  in  the  dilute  limit and  for 
hypothetical  zinc-blende  MnAs  using pseudoatomic-orbital-based  LDA
algorithms  \cite{sanvitojs02}.  The  calculated  lattice constant  of
MnAs is similar to the  lattice constant of GaAs.  This result 
appears to suggest that the relatively large compressive
strains observed  in (Ga,Mn)As epilayers  grown on GaAs are  caused by
intrinsic defects  rather than  by the substitutional Mn.  

\vspace{-.5cm}

\begin{figure}[h]
\epsfxsize=2.3in
\hspace*{-0.7cm}
\rotatebox{-90}{\centerline{\epsffile{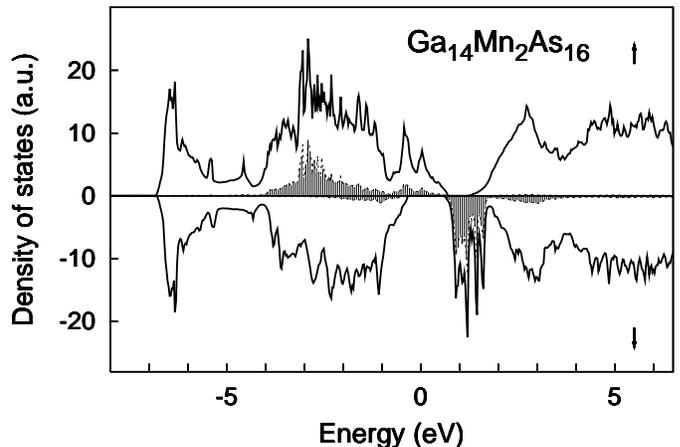}}}
\vspace{-1.3cm}
\caption{LDA Spin-polarized density of states of Ga$_{14}$Mn$_{2}$As$_{16}$
crystal with substitutional Mn atoms \protect\cite{macaprb02}. 
Grey areas represent the contribution 
from Mn 3d states. Energy is measured from the Fermi level.   
}
\label{masek}
\end{figure}

The  electronic  structure of  (Ga,Mn)As  supercells  with  Mn in  the
substitutional position, with As-antisites, or Mn-interstitial defects
have  been  studied  by various methods within  the LDA   
\cite{sanvitojs02,macaprb02,erwin0209329}.   Typical spin-polarized 
band and Mn 3d projected density-of-states results are 
shown in Fig.~\ref{masek} \cite{macaprb02}
for substitutional Mn. It  has  been
conclusively established  by the LDA calculations
that in  the presence of  substitutional Mn,
the Fermi  energy  crosses  the  valence  band near  the  $\Gamma$  point,
introducing one  hole per Mn.  
Clear theoretical evidence for 
As 4p-like band  holes has been  obtained in LDA+U
calculations \cite{parkjmmm02} and the Mn  3d
density of states calculated by this method is in a 
good agreement with photoemission data 
\cite{okabayashiprb99}.  
As-antisite  and  Mn-interstitial  defects  act  as  a  double  donor,
according  to the  microscopic calculations 
\cite{sanvitojs02,macaprb02,erwin0209329}, and, therefore,  are the
likely sources of carrier compensation observed in experiment.
In summary, first-principles calculations support the picture of 
carrier induced ferromagnetism in some (III,Mn)V ferromagnets, but detailed
predictions depend on the approximation used for exchange and correlations
and also perhaps on technical details of the calculation.  This activity 
is on-going and further progress should be expected. 

\begin{figure}[h]
\epsfxsize=3.2in
\centerline{\epsffile{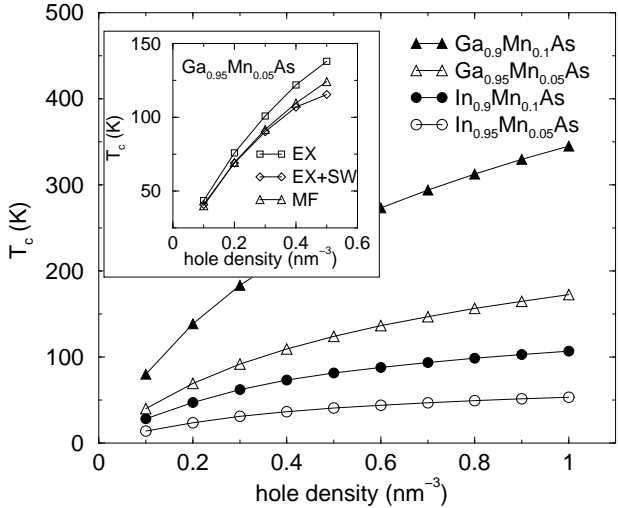}}
\caption{Main plot: Mean-field critical temperature calculated using  
the effective Hamiltonian and virtual crystal approximations for (Ga,Mn)As
(triangles) and (In,Mn)As (circles) DMSs as a
function of the hole density \protect\cite{dietlprb01,jungwirthprb02}.
Inset: Critical temperatures in (Ga,Mn)As obtained by including hole-hole
exchange interaction (EX) and spin-wave fluctuations (EX+SW) is compared
to the mean-field results (MF).}
\label{tc}
\end{figure}

Semi-phenomenological effective 
Hamiltonian  theories in which the low-energy degrees of 
freedom are Mn ion  S=5/2 spins and valence band holes,  discussed in the
following paragraphs, have been developed to make quantitative
predictions for zero and finite temperature  magnetic, transport
and  optical  properties  of  bulk samples \cite{dietlsst02,konigspringer02}  
and  heterostructures \cite{leesst02}.  In
(Ga,Mn)As, e.g., these models are partially justified by the negligible changes
in the  valence band structure of the host semiconductor  observed in
angle-resolved  photoemission  experiments \cite{okabayashiprb01}.    
The  key  term  in  the
effective  Hamiltonian   description  is  a   Kondo-like exchange interaction,
$J_{pd}\sum_I {\bf S}_I \cdot ×  {\bf s}({\bf R}_I)$,
which arises microscopically from sp-d hybridization and 
strong d-orbital Coulomb interactions.  
Here ${\bf S}_I$ is the $S=5/2$ local moment operator  on Mn site $I$ and ${\bf s}({\bf
r})$  is the  multi-band  envelope function  hole spin-operator 
\cite{dietlprb01,abolfathprb01}.   The
simplest  version of this  model combines  mean-field theory  with a 
virtual  crystal  approximation  for the  random  distribution  of  Mn
moments \cite{dietlsst02,dietlscience00,konigspringer02,jungwirthprb99}. 
Illustrative results of the model are presented in Fig.~\ref{tc} for the
critical temperature in (Ga,Mn)As and (In,Mn)As.  These calculations
and the model predictions
for the strain-engineered
magnetic anisotropy, anomalous Hall conductivity \cite{jungwirthprl02}, 
domain structure \cite{dietlkonigprb01}, and
magnetic  circular
dichroism \cite{dietlprb01} are in good agreement with experiment.

\vspace{-2cm}

The  reliability of  the  standard mean-field  approximation has  been
examined  by accounting  for the  Coulomb interactions  among 
holes \cite{konigspringer02},
which  enhance the critical  temperature, and  for correlations  in Mn
moment orientations which reduce the energetic cost of incompletely aligned spin
configurations \cite{konigprl00} 
and therefore  lower the  critical  temperature. The
calculations show  that heavy-hole - light-hole  mixing is responsible
for  a  relatively  large  spin  stiffness,  resulting  in  only  $\sim 20$~\%
reduction of the critical temperature due to spin-wave fluctuations in
arsenides  and antimonides \cite{jungwirthprb02}. 
Moreover,  the reduction  is practically
canceled  by  the enhancement  of  the  critical  temperature due  to
hole-hole interactions  (see inset of Fig.~\ref{tc})
which explains  the success of  the mean-field
theory in these ferromagnets \cite{jungwirthprb02}.

\begin{figure}[h]
\epsfxsize=3.8in

\vspace*{-.4cm}

\hspace*{-.3cm}
\centerline{\epsffile{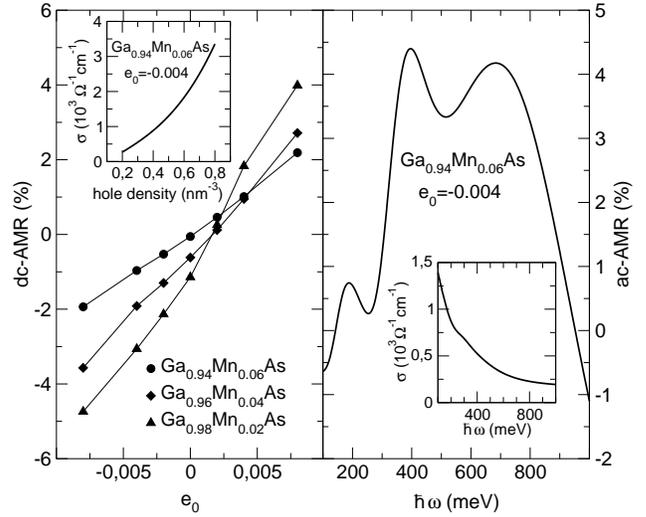}}
\vspace{-.5cm}

\caption{Left  panel: dc anisotropic magnetoresistance  (AMR) as a
function  of  strain for  50\%  compensated (due to As-antisites) 
(Ga,Mn)As  DMSs.  AMR  is
defined as the relative  difference between resistivities for in-plane
magnetization   oriented   along  the   current   direction  and   for
magnetization perpendicular to the plane of the DMS epilayer. $e_0$ is
the relative difference between the substrate and the epilayer lattice
constants.   Inset shows the  absolute conductivity  for magnetization
along  the current  direction. Right  panel: ac-AMR  as a  function of
frequency.  Inset shows  the absolute  ac-conductivity.}
\label{AMR}
\end{figure}

Theoretical models beyond the  virtual crystal approximation have been
used to  study the  effects of disorder  on transport  properties, and
magnetic properties  in general, of  the DMSs. The  Boltzmann equation
with Born  approximation scattering  rates have provided  estimates of
the anisotropic magnetoresistance effect (AMR) of order $\sim 1-10$~\%
\cite{jungwirth0206416},    in   good   agreement    with   experiment
\cite{baxterprb02,gallagher}.  AMR of a similar magnitude is predicted
by   Kubo  formula   for  the   ac-conductivity,  as   illustrated  in
Fig.~\ref{AMR}.  The key for  understanding the transport and magnetic
anisotropy  effects  is  a  strong  spin-orbit coupling  in  the  host
semiconductor  valence-band.  The  absolute dc-conductivity  estimates
(see  the left  inset of  Fig.~\ref{AMR})  \cite{jungwirth0206416} are
typically   only   2-10   times   larger  than   experimental   values
\cite{baxterprb02,edmonds0205517,potashnikprb02}.   The discrepancy is
likely partly  due to inaccuracies in the  model scattering amplitudes
and multiple-scattering or localization effects omitted
by this approximation.   However experimental conductivities are still
increasing as growth and  annealing procedures are optimized 
so the discrepancy
could be  due mainly to  unintended extrinsic disorder.   Kubo formula
calculations  of  infrared conductivities,  illustrated  in the  right
inset of  Fig.~\ref{AMR}, of metallic (Ga,Mn)As  DMSs demonstrate that
transverse  f-sum rule measurements  can be  used to  extract accurate
values   for  the  free   carrier  density   \cite{sinovaprb02}.   The
hole-fluid effective  Hamiltonian theory discussed above  uses no free
parameters  and  is expected  to  be  reliable  in samples  with  high
critical  temperatures   where  the  holes  are   metallic  and  their
interaction with Mn and other impurities is effectively screened.

More detailed studies of disorder  effects in DMSs have combined Kondo
description  of  the spin  interactions  with  Monte Carlo  techniques
applied to both metallic \cite{schliemannprb02} 
(high carrier density) and insulating \cite{bhattjs02} (low
carrier density) regimes, and with the CPA \cite{bouzerar0208596} 
or spin-wave theory \cite{schliemannprl02} for
the  metallic  samples.   
The CPA  calculations, which
start  from a  tight-binding  approximation band  model  and use  RKKY
theory  of the  effective Mn-Mn  interaction,  predict a  factor of  3
enhancement of the critical temperature due to disorder for typical Mn
and hole densities in (Ga,Mn)As  DMSs \cite{bouzerar0208596}. 
Other studies using dynamical mean field theory \cite{millisprl01}
suggest similar trends with some additional features when considering
the effective $J_{pd}$ coupling as a variable.
On the other hand, the random
Mn distribution and spatial fluctuations of the magnetization suppress
the    ferromagnetic   transition    temperature   in    Monte   Carlo
simulations \cite{schliemannprb02}.  
The non-collinear nature  of the  ground state  due to
disorder has been pointed  out in the spin-wave theory calculations
\cite{schliemannprl02}.
It was  also suggested \cite{zerandprl02} 
that highly  frustrated magnetic correlations
may result from the strong  spin-orbit interaction in the valence band
and the spatial disorder.

The  metal-insulator transition  regime  was
explored by exact  diagonalization techniques \cite{timmprl02,yang0202021}.   
Consistent  with experiment, the studies  predict a
metal-insulator transition for 1\% Mn  doping in GaAs DMSs 
\cite{timmprl02,yang0202021}. It has been
emphasized \cite{timmprl02} 
that  defect correlations play  an important role  in the
transition and  the calculations \cite{timmprl02,yang0202021} 
have provided evidence  in favor of
the  applicability of  hole-fluid  and impurity-band  magnetic-polaron
models   on  metallic   and   insulating  sides   of  the   transition
respectively. Monte  Carlo studies of the insulating  phase have found
spatially   inhomogeneous  magnetizations  bellow   the  ferromagnetic
critical temperature  and an  enhancement of the  critical temperature
due to disorder \cite{bhattjs02}.

We acknowledge helpful discussions with Tomasz Dietl, 
and Jan Ma\v{s}ek who also provided  figures presented in this
review. The work was supported by the Grant Agency of the Czech
Republic, DARPA, DoE, and the Welch Foundation.

\end{document}